\documentclass{aastex}
\usepackage{graphicx,emulateapj5,natbib}
\def\myputfigure#1#2#3#4#5%
{\vskip#5pt\makebox[0pt]{\hskip#2in
\includegraphics[width=#3\textwidth]{#1}}\vskip#4pt\hfill}

\def\beb{\begin{thebibliography}{999}}
\def\eeb{\end{thebibliography}}

\def\be{\begin{equation}}
\def\ee{\end{equation}}
\def\bea{\begin{eqnarray}}
\def\eea{\end{eqnarray}}
\def\o{\over}

\begin{document}

\shorttitle{Cluster Evolution, Surveys and Dark Energy}
\shortauthors{Majumdar \& Mohr}
\submitted{submitted to ApJ June 14, 2002}

\title{Importance of Cluster Structural Evolution in Using X--ray and SZE Galaxy Cluster Surveys to Study Dark Energy}
\author{Subhabrata Majumdar\altaffilmark{1}  \& Joseph J Mohr\altaffilmark{1,2}}
\altaffiltext{1}{Department of Astronomy, University Of Illinois, 1002 West Green St., Urbana, IL 61801}
\altaffiltext{2}{Department of Physics, University Of Illinois, 1002 West Green St., Urbana, IL 61801}
\email{subha@uiuc.edu}\email{jmohr@uiuc.edu}

\begin{abstract}
We examine the prospects for measuring the dark energy equation of state parameter $w$ within the context of the still uncertain redshift evolution of galaxy cluster structure.   We show that for a particular X--ray survey (SZE survey) the constraints on $w$ degrade by roughly a factor of 3  (factor of 2) when one accounts for the possibility of non--standard cluster evolution.  With followup measurements of a cosmology independent, mass--like quantity it is possible to measure cluster evolution, improving constraints on cosmological parameters (like $w ~\& ~\Omega_M$).  We examine scenarios where 1\%, 10\% and 100\% of detected clusters are followed up, showing that even a modest followup program can enhance the final cosmological constraints.   For the case of followup measurements on 1\% of the cluster sample with an uncertainty of $30\%$ on individual cluster mass--like quantities,  constraints on $w$ are improved by a factor of 2 to 3. For the best case scenario of a zero curvature universe,  these particular X--ray and SZE surveys can deliver uncertainties on $w$ of $\sim$4\% to 6\%.
\end{abstract}


\section{Introduction}

Galaxy clusters have been used extensively to determine the cosmological matter density parameter and the amplitude of density fluctuations.   Cluster surveys in the local universe are particularly useful for constraining a combination of the matter density parameter $\Omega_M$ and the normalization of the power spectrum of density fluctuations \citep[we describe the normalization using $\sigma_8$, the {\it rms} fluctuations of overdensity within spheres of 8$h^{-1}$~Mpc radius; i.e.][]{henry97,viana99, reiprich02}; surveys that probe the cluster population at higher redshift are sensitive to the growth of density fluctuations, allowing one to break the $\Omega_M$-$\sigma_8$ degeneracy that arises from local cluster abundance constraints \citep{eke96,bahcall98}.      \citet{wang98} argued that a measurement of the changes of cluster abundance with redshift would provide constraints on the dark energy equation of state parameter $w\equiv p/\rho$. 

Describing the problem in terms of cluster abundance only makes sense in the local universe , because, of course, one cannot measure the cluster abundance without knowing the survey volume; the survey volume beyond $z\sim0.1$ is sensitive to cosmological parameters that affect the expansion history of the universe-- namely, the matter density $\Omega_M$, the dark energy density $\Omega_E$ and the dark energy equation of state $w$.  A cluster survey of a particular piece of the sky with appropriate followup actually delivers a list of clusters with mass estimates and redshifts-- that is, the redshift distribution of galaxy clusters above some detection limit.

Recently, it has been recognized that with current instrumentation it is possible to use such surveys of galaxy clusters extending to redshifts $z>1$ to precisely study the amount and nature of the dark energy \citep{haiman01}.  Clusters are promising tools for precision cosmological measurements, because they exhibit striking regularity and they exist throughout the epoch of dark energy domination.  Moreover, their use is complementary to studies of cosmic microwave background (CMB) anisotropy and SNe Ia distance measurements \citep{haiman01,levine02,hu02}.  Following
\citet{haiman01}, a series of analyses appeared that explore the theoretical and observational obstacles to precise cosmological measurements with cluster surveys : \citet{holder01b} applied the Fisher matrix formalism to the cluster survey problem and showed that high yield SZE cluster surveys can provide precise constraints on the geometry of the universe through simultaneous measurements of $\Omega_E$ and $\Omega_M$.  \citet{weller01} demonstrated that future SZE surveys might constrain the variation of the dark energy equation of state $w(z)$.  \citet{hu02} examined the effects of cosmic variance on cluster surveys as well as including the effects of imprecise knowledge of a more complete list of cosmological parameters. \citet{levine02} examined an X--ray cluster survey, showing that a sufficiently large survey allows one to measure cosmological parameters and constrain the all--important cluster mass--observable relation simultaneously.

An important caveat to these works is that the authors assumed that the evolution of cluster structure with redshift was perfectly known.
In this paper, we examine the effects of uncertainties about cluster structural evolution on cosmological constraints from cluster surveys, finding that current survey projections that ignore this evolution uncertainty overstate the cosmological sensitivity of the survey.  Furthermore, we examine the effects of survey followup to measure a cluster mass--like quantity $M_f$, demonstrating that an appropriately designed survey can overcome this evolution uncertainty.  

In addition, our calculations underscore the importance of incorporating information from multiple observables into future cluster surveys.
Clusters of galaxies are dark matter dominated objects with baryon reservoirs in the form of an intracluster medum (ICM) and a galaxy population.  Clusters can be found through the light the galaxies emit, the gravitational lensing distortions the cluster mass introduces into the morphologies of background galaxies, the X--rays emitted by the energetic ICM, the distortion that the hot ICM introduces into the cosmic microwave background spectrum (SZE), and the effects that the ICM has on jet structures associated with active galaxies in the cluster.  These methods are largely complementary, each having different strengths.  It appears that X--ray and SZE signatures of clusters are higher contrast observables than are weak lensing or galaxy light.  That is, massive galaxy clusters are more prominent relative to the far more abundant lower mass halos and the large scale filaments when viewed with the SZE and X-ray;  projection effects are a far more serious concern when using galaxy light or weak lensing signatures.  Studies of the highest redshift galaxy clusters will likely be done with the SZE, because of the redshift independence of the spectral distortion in the CMB.  Any effort to carry out a precise cosmological study using galaxy clusters will undoubtedly be most effective through some combination of these complementary, cluster observables.

The paper is arranged in the following way.  In $\S$\ref{sec:surveys} we describe two representative surveys and survey followup.  Section~\ref{sec:fisher} contains a description of our estimates of the survey sensitivity when followup is included as well as a description of our fiducial model.  Results are presented in $\S$\ref{sec:results} and discussed further in $\S$\ref{sec:conclusions}.

\section{Future Galaxy Cluster Surveys}
\label{sec:surveys}
A study of using cluster surveys to probe dark energy begins with the redshift distribution of detectable clusters within a survey solid angle $\Delta\Omega$,
\be
{{dN}\o{dz}} ~ = ~ \Delta\Omega {{dV}\o{dzd\Omega}}(z)\int_{0}^\infty f(M) {{dn(M,z)}\o{dM}}dM
\ee
where ${{dV}/{dzd\Omega}}$ is the comoving volume element, $({{dn}/{dM}})dM$ is the comoving density of clusters of mass $M$, and $f(M)$ is the cluster selection function for the survey.  In this analysis we take $f(M)$ to be a step function at some limiting mass $M_{lim}$, which corresponds to the mass of a cluster that lies at the survey detection threshold.  We use the cluster mass function ${{dn}/{dM}}$ determined from structure formation simulations \citep{jenkins01}.

In practice surveys select clusters using observables like the X--ray flux, SZE flux, galaxy light or weak lensing shear.  Thus, in addition to the ingredients above, one requires a virial mass--observable relation (like $M$--$L_x$, $M$--$L_{sz}$ or $M$--$\gamma_t$). 
Low redshift clusters do exhibit regularity \citep[i.e.][]{david93,mohr97a}, suggesting that observables like the ICM X-ray luminosity and temperature are good mass estimators \citep{finoguenov01,reiprich02}.  Cluster mass to light ratios have been studied for decades, and it may be that this body of work together with modern datasets will allow more conclusive statements about how well galaxy light traces cluster halo mass.  Hydrodynamical simulations lead us to expect that the SZE luminosity (related to the total thermal energy within the virial region) should be the best ICM observable for predicting mass, but we await new observations with next generation SZE instruments to demonstrate this.

A central feature of these mass--observable relations is that they evolve with redshift due to the increasing density of the universe at earlier times (and the changing ratio of distance to lense and source in the case of weak lensing).  Within standard structure formation models, galaxy clusters form self--similarly, and so there are standard evolution models for each mass--observable relation \citep[e.g.][]{bryan98,mohr00a,evrard02}.  
Results to date suggest that the degree of cluster regularity locally and at intermediate redshift is comparable \citep{mushotzky97,mohr00a,vikhlinin02}.
However, given the central importance of cluster mass estimates in using surveys to study dark energy, we can only regard these standard structure formation models as a guide;  ultimately, one needs to determine the evolution of cluster structure observationally.  In this paper we examine the effects that non--standard redshift evolution of cluster structure would have on our ability to use cluster surveys to study the dark energy.  As detailed below, we explore non--standard redshift evolution by allowing an additional dependence of $(1+z)^\gamma$ in the evolution of the relevant mass--observable relations.

\subsection{An X--ray and an SZE Survey}

To examine these new effects, we adopt two representative surveys that are being promoted as ways of measuring the dark energy equation of state.  Namely, we examine the following two high yield surveys:  (i) a $10^4$~deg$^2$ flux limited X--ray survey proposed as part of the DUET mission to the NASA Medium--class Explorer Program, and (ii) a 4,000~deg$^2$ SZE survey to be carried out with a proposed 8m South Pole Telescope (SPT).  Figure~\ref{fig:surveys} contains a plot of the redshift distribution and limiting mass for both surveys.

We model the DUET X-ray survey as having a bolometric flux limit of $f_x>1.25\times10^{-13}$~erg/s/cm$^2$ (corresponding to $f_x>5\times10^{-14}$~erg/s/cm$^2$ in the 0.5:2~keV band).  For our fiducial cosmological model (see $\S$\ref{sec:fiducial} below) this survey yields $\sim$21,600 detected clusters, consistent with the known X-ray $\log N$--$\log S$ relation for clusters \citep[e.g.][]{gioia01}.  
For our mass--observable relation, we adopt a bolometric X-ray luminosity--mass relation
\be
f_x(z)4\pi d_L^2 = A_x M_{200}^{\beta_x} E^2(z)\left(1+z\right)^{\gamma_x}
\label{eq:lx-m}
\ee
where $f_x$ is the observed flux in units of erg~s$^{-1}$cm$^{-2}$, $d_L$ is in units of Mpc, $M_{200}$  in units of $10^{15}M_\odot$ is the mass enclosed within a radius $r_{200}$ having a overdensity of $200$  with respect to critical and  $H(z)=H_0E(z)$. 
where $E^2(z)=\Omega_M(1+z)^3+\Omega_k(1+z)^2+\Omega_E^{3(1+w)}$.  The $E(z)$ factors follow the evolution of the critical density of the universe $\rho_{crit}=3H^2/8\pi G$.  
We  convert $M_{200}$ to $M(z)$, the halo mass appropriate for our mass function at redshift $z$ using a halo model \citep[][hereafter, NFW; see discussion below regarding the effects of uncertainties in this conversion]{navarro97}. 
Our standard evolution model ignores the $T^{1/2}$ dependence of the bolometric bremsstrahlung radiation, because X--ray surveys detect clusters using detected photons rather than detected energy.
We introduce the possibility of non-standard evolution of the mass--observable relation with the parameter $\gamma_x$.   
We take $\gamma_x=0$ to be consistent with the observed weak evolution in the luminosity--temperature relation \citep{vikhlinin02}, and we choose $\beta_x=1.807$ and $\log{A_x}=-3.926$, consistent with observations \citep{reiprich02}.  The overall $h$ scaling of the limiting mass is $h^{-1.11}$. 

We model the SPT SZE survey as a flux limited survey with $f_{SZ}>5$~mJy at 150~GHz.  Within our fiducial cosmological model this survey would yield $\sim 13,500$ clusters with measured fluxes.  The mass--observable relation is
\bea
\label{eq:lsz-m}
f_{sz}(z,\nu) d_A^2 &=&3.781 {f(\nu)f_{ICM}}T{{M_{200}}}\left(1+z\right)^{\gamma_{sz}}\nonumber\\
M_{200} &=& A_{sz} {{\left(k_BT\right)^{\beta_{sz}}} \o{E(z)}}
\eea 
where $f(\nu)$ is the frequency dependence of the SZE distortion, $f_{sz}$ is the observed flux in mJy, $T$ is in Kelvin, $M_{200}$ is in units of $10^{15}M_\odot$, $f_{ICM}=0.12$ \citep[e.g][]{mohr99} and $d_A$ is in units of Mpc  \citep[see][]{diego02}.    We use $\log{A_{sz}}=13.466$, $\beta_{sz}=1.48$ \citep{finoguenov01} and $\gamma_{sz}=0$ to model standard structure evolution. In this form, the overall $h$ scaling of the limiting mass is $h^{-1.61}$.  Note that in determining the estimated uncertainties on cosmological parameters, we allow the normalization of these mass--observable relations to be free to vary.  The survey contains enough information to solve for the best normalization and the cosmological parameters simultaneously;  therefore, shifts in model inputs like $f_{ICM}$ within the observational uncertainties have minimal effect on our conclusions.

\myputfigure{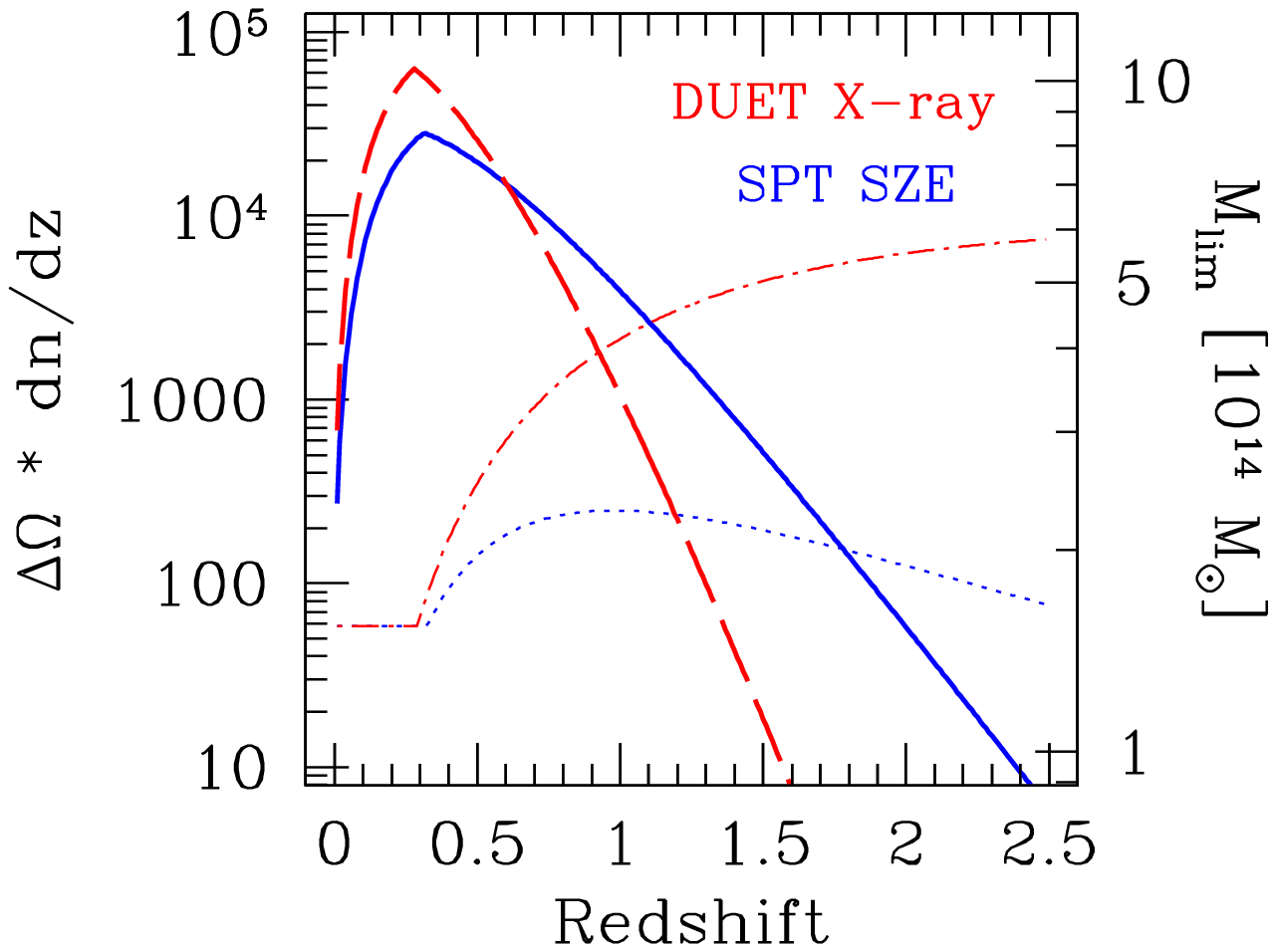}{3.2}{0.5}{-20}{0}
\figcaption{The cluster redshift distribution (heavy line) and mass limit (light line) of the $10^4$~deg$^2$ DUET X--ray survey (dashed) and the $4,000$~deg$^2$ South Pole Telescope (SPT) SZE survey (solid).  The surveys are flux limited ($f_x>1.25\times10^{-13}$erg/s/cm$^2$ and $f_{sz}>5$mJy at 150~GHz), and we impose a minimum cluster mass of $1.54\times10^{14}$~$M_\odot$.
\label{fig:surveys}\vskip5pt}

A generic problem with flux limited surveys is that at low redshift the implied mass limit drops well below those masses corresponding to galaxy clusters.  The flux from a nearby object is spread over a much larger portion of the sky, and surface brightness selection effects become important.  We model these complications by imposing a minimum cluster mass of $10^{14}h^{-1}$~M$_\odot$.  This lower limit on the survey mass limit is readily apparent below $z \sim 0.25$ in Fig~\ref{fig:surveys}.

\subsection{Followup of Large Solid Angle Surveys}
The redshift distribution of clusters contains far more cosmological information than the surface density of clusters \citep{haiman01} or the angular correlation function \citep[e.g.][]{komatsu02}.  Thus, in both these surveys each detected cluster will be followed up with multi--band optical and near-IR photometry to provide photometric redshift estimates.  These same data can be used to estimate cluster masses through their weak lensing effects on background galaxies \citep[e.g.][]{bartelmann01b} and the total detected light from cluster galaxies.

In addition,  some of these clusters can be followed up with detailed X--ray, SZE or galaxy spectroscopic observations that allow one to measure the mass-like quantity $M_f(\theta)=M(\theta)/d_A$, which we will refer to as the followup mass.  As an example, in the case of followup X--ray observations that deliver the projected ICM temperature profile and surface brightness profile, it is straightforward to extract the underlying ICM density $\rho(d_A\theta)$ and temperature profile $T(d_A\theta)$ to then estimate the followup mass $M_f$ as
\begin{equation}
M_f(\theta)=-\theta {k_BT(\theta)\over G\mu m_p}\left({d\ln\rho\over d\ln\theta} + {d\ln T\over d\ln\theta}\right)
\end{equation}
where $m_p$ is the proton mass, $k_B$ is Boltzmann constant, $G$ is Newton constant, and the ICM number density $n\equiv\rho/\mu m_p$.  Note that only the {\it shape} of the ICM density and temperature profiles is required (i.e. knowledge of the actual distance to the system is not required).   The followup--mass $M_f$ can then be examined at within some angle $\theta$ along with the cluster X-ray or SZE flux.   At fixed redshift, this followup would produce an $M_f$--$f_x$ of $M_f$--$f_{sz}$ relation which would provide direct constraints on the structural evolution of the clusters.
As we will see, the parameter sensitivity of these scaling relation observations can exhibit quite different degeneracies than for the cluster redshift distribution, making the scaling relations and the cluster redshift distribution complementary.  In $\S$\ref{sec:fisher} below, we describe how these survey followup observations are included in our estimates of the cosmological sensitivity of the survey.

\begin{table*}[htb]\small
\caption{Estimated Parameter Constraints\label{tab:params}}
\hbox to \hsize{\hfil\begin{tabular}{lrrrrrrcrrr}
Description 		& \multicolumn{1}{c}{$\Omega_M$} & \multicolumn{1}{c}{$\Omega_{tot}$} & \multicolumn{1}{c}{$\sigma_8$} & \multicolumn{1}{c}{$w$} & \multicolumn{1}{c}{$h$} & \multicolumn{1}{c}{$n$} & \multicolumn{1}{c}{$\Omega_B$} & \multicolumn{1}{c}{$\log{A}$} & \multicolumn{1}{c}{$\beta$} & \multicolumn{1}{c}{$\gamma$} \\
\hline
\emph{Priors} 		&  		& 0.0100 	& 	        &                 &    0.0323 &    0.0500 & 0.0040\\
\multicolumn{10}{l}{\emph{SPT SZE Survey}}\\
\hspace{5pt}Std Evolution		& 0.0177 & 0.0077 & 0.0134 & 0.1629 & 0.0323 & 0.0478 & 0.0040 & 0.1392 & 0.0064 &      -      \\
\hspace{5pt}Non-Std Evolution 	& 0.0184 & 0.0079 & 0.0195 & 0.2487 & 0.0323 & 0.0479 & 0.0040 & 0.1505 & 0.0064 & 0.4602\\
\hspace{5pt}+ 1\% Followup 	         & 0.0167 & 0.0074 & 0.0157 & 0.1404 & 0.0285 & 0.0476 & 0.0040 & 0.1423 & 0.0064 & 0.2248\\
\hspace{5pt}+ 10\% Followup 	         & 0.0147 & 0.0070 & 0.0122 & 0.1237 & 0.0259 & 0.0472 & 0.0040 & 0.1320 & 0.0061 & 0.0958\\
\hspace{5pt}+ 100\% Followup 	& 0.0139 & 0.0068 & 0.0109 & 0.1207 & 0.0248 & 0.0471 & 0.0040 & 0.1004 & 0.0046 & 0.0353\\
\hspace{5pt}+ Flat + 100\% F-up 	& 0.0139 &      -       & 0.0109 & 0.0406 & 0.0235 & 0.0439 & 0.0039 & 0.1001 & 0.0046 & 0.0352\\
\hspace{5pt}100\% F-up Only		& 0.2360 & 0.0099 &      -       & 0.6369 &       -       &      -       &       -       & 0.2204 & 0.0067 & 0.1912\\

\multicolumn{10}{l}{\emph{DUET X--ray Survey}}\\
\hspace{5pt}Std Evolution		& 0.0147 & 0.0087 & 0.0113 & 0.1659 & 0.0323 & 0.0476 & 0.0040 & 0.0928 & 0.0060 &      -      \\
\hspace{5pt}Non-Std Evolution 	& 0.0255 & 0.0092 & 0.0441 & 0.4505 & 0.0324 & 0.0476 & 0.0040 & 0.2216 & 0.0096 & 1.2955\\
\hspace{5pt}+ 1\% Followup 		& 0.0157 & 0.0086 & 0.0167 & 0.1655 & 0.0233 & 0.0475 & 0.0040 & 0.1060 & 0.0062 & 0.3310\\
\hspace{5pt}+ 10\% Followup 		& 0.0139 & 0.0085 & 0.0116 & 0.1454 & 0.0201 & 0.0475 & 0.0040 & 0.0814 & 0.0055 & 0.1185\\
\hspace{5pt}+ 100\% Followup 	& 0.0134 & 0.0084 & 0.0105 & 0.1414 & 0.0191 & 0.0473 & 0.0040 & 0.0586 & 0.0040 & 0.0444\\
\hspace{5pt}+ Flat + 100\% F-up 	& 0.0133 &      -       & 0.0100 & 0.0593 & 0.0187 & 0.0467 & 0.0039 & 0.0585 & 0.0040 & 0.0444\\ \hspace{5pt}100\% F-up Only		& 0.2018 & 0.0100 &      -       & 0.5500 &      -       &      -        &      -       & 0.1408 & 0.0056 & 0.1899\\
\end{tabular}\hfil}
\vskip-20pt
\end{table*}

\section{Cosmological Sensitivity of a Survey}
\subsection{Fisher Matrix Technique}
\label{sec:fisher}
  We employ the Fisher matrix technique to probe the relative sensitivities of two cluster surveys to different cosmological and cluster structural parameters. The Fisher matrix information for a data set \citep[see][]{tegmark97,eisenstein98b} is defined as 
$F_{ij} \equiv <{{\partial^2 \ln{\mathcal{L}}}\o{\partial p_i \partial p_j}}>$, where $\mathcal{L}$ is the likelihood for an observable (${{dN}\o{dz}}$ for the survey and $M_f $ for the followup) and $p_i$ describes our parameter set. The inverse $F_{ij}^{-1}$ describes the best attainable covariance matrix $[C_{ij}]$ for measurement of the parameters considered. The diagonal terms in $[C_{ij}]$ then give the uncertainties on each of our parameters.  In calculating these uncertainties, we have added the Fisher matrix for the followup observations ($F_{ij,}^{f}$), the Fisher matrix for the cluster redshift distribution ($F_{ij}^{s}$) and several external  priors that will be discussed below.

We construct the survey Fisher matrix $F_{ij}^{s}$ following \citet{holder01b} as
\be
F_{ij}^{s} = \Sigma_n {{\partial dN/dz}\o{\partial p_i}} {{\partial dN/dz}\o{\partial p_j}}{{1}\o{{dN/dz}}},
\ee
where we sum over $n$ redshift bins of size $\Delta z = 0.01$ to $z_{max}=3.0$ and $dN/dz$ represents the number of surveyed clusters in each redshift bin. The Fisher matrix for the followup is constructed  as
\be
F_{ij}^{f} = \Sigma_n {dV\o dz}\int dM\, f {dn \o dM}\left({{\partial M_f}\o{\partial p_i}}{{\partial M_f}\o{\partial p_j}}  {{1 }\o{{\sigma^2_{M_f}}}}\right)
\label{eq:fishfollow}
\ee
where $M_f$ is a function of halo mass $M$ and angular radius $\theta$, and $f(dn/dM)$ represents the number of clusters of mass $M$ for which followup mass measurements are available in a particular redshift bin.  We examine cases where the followup fraction $f$ is 1\%, 10\% and 100\%.  

To generate the followup Fisher matrix, we calculate the cluster binding mass within radius $r=d_A\theta$ for a halo with virial mass $M$.  To do this calculation we assume cluster mass profiles are well represented, on average, by NFW models with concentration index $c=5$.  In practice clusters undergo merging quite frequently and there is a range of halo shapes.  This introduces a ``theoretica''l uncertainty to the followup mass.  In this analysis we take $\sigma_{M_f} = 0.3M_f$ to be the characteristic uncertainty in the followup mass measurments.  This uncertainty reflects the observational uncertainty on individual cluster followup mass measurements as well as the uncertainties inherent in predicting the followup mass from the halo virial mass.  As is clear from Eqn~\ref{eq:fishfollow}, the redshift and mass distributions of the followup clusters match those of the full cluster survey sample; that is, we don't choose to followup only high redshift clusters, which would presumably provide the tightest constraints on our evolution parameter.   We also choose $\theta$ to be a dynamically varying quantity fixed to be $95\%$ of the virial radius corresponding to the cluster limiting mass at each redshift.  Thus, followup at all redshifts corresponds to mass--like measurements at radii within the virial radius ($r_\theta(M) < r_{200}(M)$).

\subsection{Fiducial Cosmology and External Constraints}
\label{sec:fiducial}
The fiducial cosmological parameters of our model are $h=0.65$ \citep[i.e.][]{hendry01,ajhar01,reese02},  $\Omega_M=0.3$ \citep[i.e.][]{mohr99,grego01}, $\Omega_{tot}=\Omega_M+\Omega_{E}=1$ \citep[i.e.][]{netterfield01,pryke01}, $w=-1$, $n=0.96$ \citep{netterfield01}, $\Omega_B=0.047$ \citep{burles98}, and a COBE normalized $\sigma_8=0.72$ \citep{bunn97}.  Note that we use a rather low value of $\sigma_8$, which is consistent with the recent 2dF analysis \citep{lahav02}.  Because the expected number density of clusters is very sensitive to the value of $\sigma_8$, our fiducial SZE survey has fewer clusters when compared to some previous studies \citep[i.e.][]{holder01b}.

Cosmological constraints from cluster surveys are complementary to constraints from SNe Ia distance measurements and observations of the anisotropy of cosmic microwave background.  This is particularly true when it comes to using cluster surveys to measure the dark energy equation of state parameter $w$ \citep{haiman01}.  In combination with precise CMB constraints on the curvature ($\Omega_k=0$), cluster surveys enable precise measurements of the dark energy equation of state;  however, when curvature is allowed to depart from zero-- even slightly-- the cluster constraints on $w$ weaken considerably.  
For a flat universe , the constraints from our SZE survey assuming standard evolution and a $100\%$ followup  on $w$ ($\Omega_M$) are 0.0406 (0.0139), whereas for $\sigma_k=0.01$ the constraints are 0.1207 (0.0139).

For the analysis presented here, we adopt relatively conservative priors from future CMB anisotropy studies and distance measurements.  We assume the power spectrum index $n$ will be known to $5\%$ , i.e. $\sigma_n=0.05$, the Hubble parameter will also be known to $5\%$, i.e $\sigma_h=0.0325$ and the total density parameter $\Omega_{tot}$ will be known to $1\%$, i.e $\sigma_k=0.01$.  In addition, we take the prior on the baryon density parameter ($\Omega_B=0.047$) to be $\sigma_{\Omega_B}=0.004$.  For reasonable values of $\Omega_B$, surveys are affected by $\Omega_B$  variations only through minor effects on the transfer function for density perturbations \citep[see also][]{levine02}.   Finally, we neglect the possibility of a variation with redshift in the equation of state parameter $w$ \citep{weller01}.

\section{Cosmological Parameter Constraints}
\label{sec:results}
Our results are listed in Table~1 and highlighted in the following figures.
Table 1 contains a listing of 1$\sigma$ uncertainties on all seven cosmological and three mass-observable relation parameters (see Eqns~\ref{eq:lx-m} \& \ref{eq:lsz-m} for definitions).  The first line contains a listing of the priors adopted for the runs.   Following that are the results for the SZE survey and then the X-ray survey.    For each survey we show the constraints in the case of standard evolution (i.e. $\gamma=0$) followed by non--standard evolution ($\gamma$ is free parameter).  The lines that follow highlight the impact of 1\%, 10\% and 100\% followup.  Following those scenarios is what we consider to be the ideal case of a flat universe with 100\% survey followup.  Finally, we show the case for followup only (i.e. the Fisher matrix derived from the redshift distribution isn't used).  The constraints in this line provide some insights into the parameter leverage that is afforded by cluster followup observations.  In all cases the uncertainties are absolute (i.e. $\sigma_{M}=0.0177$ means $\Omega_M=0.3\pm0.0177$).

\myputfigure{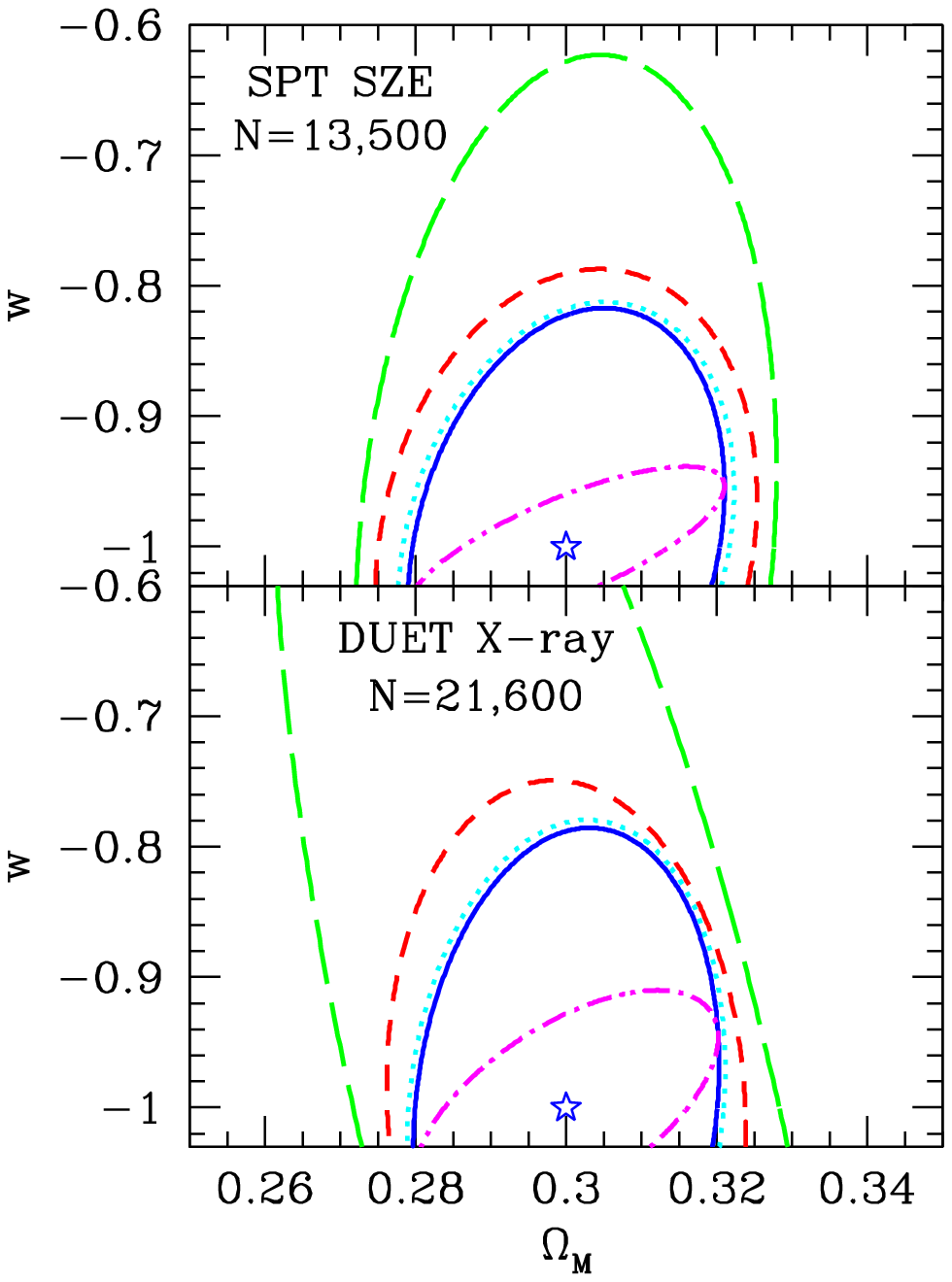}{3.2}{0.5}{-20}{-0}
\figcaption{Constraints on $w$ and $\Omega_M$  for an SZE (above) and 
an X-ray survey (below).  The star marks the fiducial model.  Contours denote joint 
1$\sigma$ constraints in five scenarios: constraints from $dN/dz$ where (i--long dashed) 
the cluster evolution is unknown;  
constraints from $dN/dz$ and followup mass measurements for (ii--short dashed) 1\% 
of sample, (iii--dotted) 10\% of sample, and (iv--solid) 100\% of sample. The case for
100\% followup plus added prior of a flat universe is also shown (v--dot--dashed). 
The  followup mass measurements are estimated to have fractional uncertainties of 
30\%.
\label{fig:cosmo}\vskip5pt}

\subsection{Importance of Non-Standard Evolution}

 In the standard evolution case, the SZE and X--ray surveys compare favorably,  yielding  1$\sigma$ absolute uncertainties on $w$ ($\Omega_M$) of  0.1629 and 0.1659 (0.0177 and 0.0147), respectively.  However, when one takes into account the possibility of non-standard evolution, the constraints on $w$ weaken by almost a factor of  2  to $\sim0.25$ for SZE and a factor 3 to $\sim0.45$ for X--ray; $\Omega_M$ constraints weaken by close to a factor of 2  to $\sim 0.026$ for the X--ray but are only slightly affected in the SZE survey.  The constraints from the cluster redshift distribution $dN/dz$ on $\gamma_{sz/x}$ are very weak at 0.46 and 1.29, respectively; this large uncertainty in the evolution of the mass--observable relation contributes to the weakened sensitivity to other cosmological parameters.  

The importance of evolution in interpreting the cluster redshift distribution contrasts somewhat with the results of the \citet{levine02} study, which showed that prior knowledge of the normalization of the mass--observable relation has only a weak effect on the cosmological sensitivity of cluster surveys \citep[see also][]{diego01}.  In their study, they only considered the standard evolution model.  Within the context of uncertain evolution of the mass--observable relation one needs observations in addition to $dN/dz$ to determine the evolution parameter $\gamma$ and regain high sensitivity to the equation of state parameter $w$.  Next we examine the effects of including followup mass measurements.

\subsection{Effects of Followup Mass Measurements}
Figure \ref{fig:cosmo} contains joint constraints on $\Omega_m$ and $w$ for the two surveys that highlight the effect of survey followup.  For each survey we show constraints with non-standard  evolution and no followup (long dashed), 1\% (dashed), 10\% (dotted), and 100\% (solid)  followup  along with 100\% followup in a flat universe (dot--dashed).  The figure makes clear that even a limited followup program can greatly improve cosmological constraints.   

Table 1 shows that as the followup fraction increases, the constraints on $w$ in the SZE survey tighten from 0.25 (no followup) to 0.14 (1\% followup) to 0.12 (10\% and 100\% followup).
The difference between 10\% and 100\% followup is minimal, suggesting that the ten times higher cost of full survey followup is not a worthwhile investment when viewed solely from the perspective of obtaining constraints on the equation of state of the dark energy.  

Notice that the constraints on $w$ in the cases of even limited followup are somewhat better than the constraints in the cases where we assume complete knowledge of the evolution of the mass--observable relation.  In the X-ray survey, a program to followup as few as 1\% of the clusters can offset the increase in uncertainties that we see in going from standard evolution to non-standard evolution. In the SZE survey, 1\% followup produces constraints that are somewhat better, reducing the uncertainty on $w$ from 0.1629 in the standard evolution scenario to 0.1404 in the non-standard evolution + 1\%followup.  This can be traced to our assumption in these calculations that the redshift distribution of followup mass measurements matches the redshift distribution of the full survey.  The higher redshift followup measurements contain more information about evolution, and the SZE survey probes to higher redshift than does the X--ray survey (see Fig.~\ref{fig:surveys})
   
Table~1 contains a listing of the effects of followup on all parameters.  It is clear that followup mass measurements dramatically reduce the projected uncertainties on cosmological and scaling relation parameters. As is evident from the last column in the table, even a modest followup of $10\%$ of the clusters reduces the uncertainty on $\gamma$ from 0.46 to 0.10 for the X-Ray case and 1.29 to 0.12 for the SZE survey.  With full followup, the constraint on $w$ shrinks from 0.25 to 0.12 in the SZE and 0.45 to 0.14 in the X--ray survey.  Even with followup of only 1\% of the clusters in the SZE survey, one reduces the uncertainty on $w$ by half.    
Comparison of the uncertainties on $\gamma$ for non-standard evolution with no followup to the  followup only case shows clearly that followup is very effective in constraining $\gamma$. For example, for the  X--Ray case the followup itself can constrain $\gamma$ to $\sim 0.19$ compared to $1.29$ that one can achieve from the survey only. This underscores the advantage of having  a followup program  for cluster surveys. 

\myputfigure{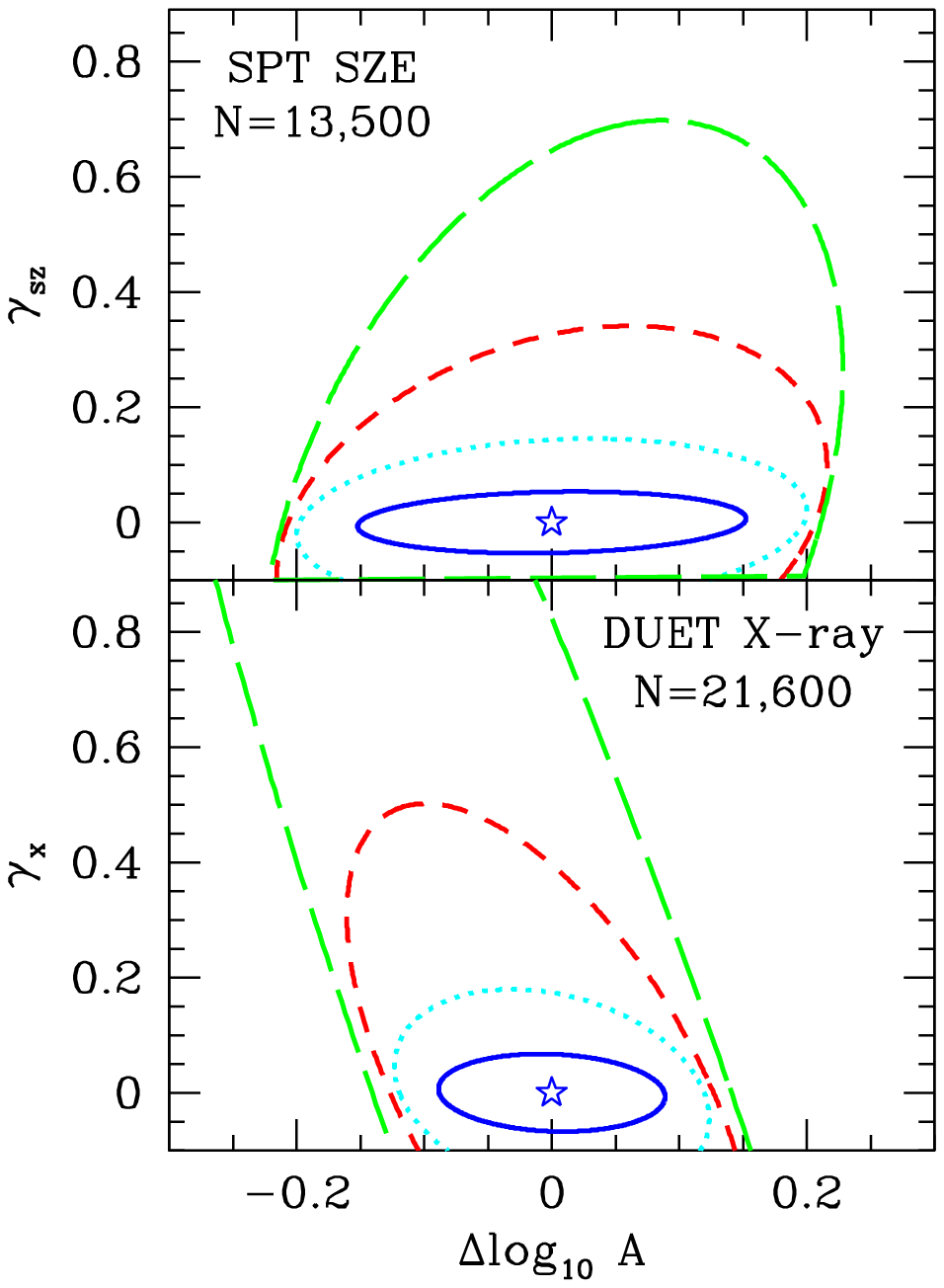}{3.2}{0.50}{-20}{0}
\figcaption{Constraints on  the mass--observable relation normalization $A$ and 
redshift evolution $(1+z)^\gamma$  for an SZE (above) and an X-ray survey (below).  
The star marks the fiducial model.  Contours denote joint 
1$\sigma$ constraints in four scenarios: constraints from $dN/dz$ where (i--long dashed) 
the non standard cluster evolution is unknown;  
constraints from $dN/dz$ and followup mass measurements for (ii--short dashed) 1\% 
of sample, (iii--dotted) 10\% of sample, and (iv--solid) 100\% of sample.  
The  followup mass measurements are estimated to have fractional uncertainties of 
30\%.
\label{fig:evolve}\vskip5pt}

In Fig. \ref{fig:evolve} we show the constraints on the mass-observable relation normalization $A$ and the evolution parameter $\gamma$ for the four cases: no followup (long--dashed) and a followup of 1\% (short--dashed), 10\% (dotted) and 100\% (solid) of the clusters.  Followup has strikingly different effects in the SZE and X-ray surveys.  Followup in the SZE survey is  more effective at constraining the evolution parameter $\gamma_{sz}$, due to the greater redshift depth of this survey.  The differences in the constraints on $\log_{10}A$ generally reflect the different definitions of the normalization and its relationship to halo mass (see Eqns~\ref{eq:lx-m} \& \ref{eq:lsz-m}).    In contrast to the previous figure that shows the effects of followup on $w$ constraints, it is clear from this figure that if one really wants to understand the normalization and evolution of the mass--observable relations, more followup is better.

\myputfigure{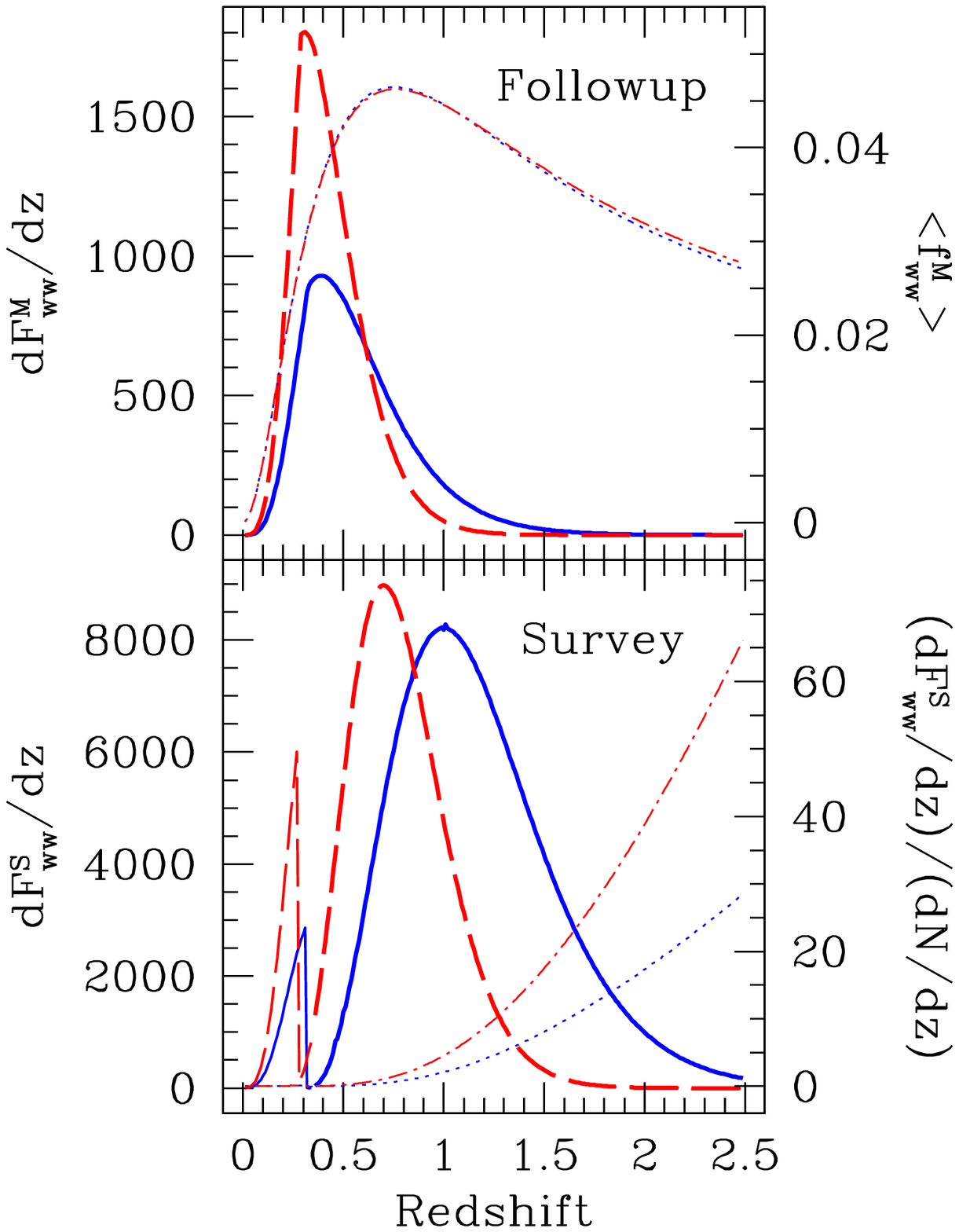}{3.25}{0.50}{-20}{0}
\figcaption{The sensitivity of the Followup (above) and Survey (below) to change in $w$
is shown for both X--Ray and SZE surveys. The four cases are: (i--solid) SZE survey
weighted sensitivity of $F_{ij}$ to change in $w$ and  
(ii--dashed) X--Ray survey weighted sensitivity $F_{ij}$ to change in $w$ (see Eqn 5 \&
Eqn 6 for an explanation of $F_{ij}$); (iii--dot-dashed) same as (i) but per unit cluster
and (iv--dotted) same as (ii) but per unit cluster. The redshift distribution of the
clusters are shown in.
\label{fig:zsensitivity}\vskip5pt}

\subsection{Redshift Variation of Parameter  Sensitivity}
 In Fig \ref{fig:zsensitivity} we display an estimate of the redshift variation of the survey and followup sensitivity to $w$.  We do this by examining the derivative with respect to redshift of the $w$-$w$ components of the survey and followup Fisher matrices.  These are shown using heavy lines for both the X-ray (dashed) and SZE (solid) surveys.  We also show an estimate of the per unit cluster sensitivity using lighter lines for the X-ray (dot--dashed) and SZE (dotted) surveys.  The axes for the heavy curves are included on the left of the figure, and the axes corresponding to the per unit cluster values are located on the right of the figure.  In both the X-ray and SZE cases and for both followup and the survey itself, the heavy curves show sensitivity that peaks at lower redshift than in the lighter curves that show the per unit cluster sensitivity.  This is simply a reflection of the redshift distributions of the clusters detected in the surveys, which peak at $z<0.5$ in both surveys.

Consider now the lower panel.  The higher redshift nature of the SPT SZE survey relative to the DUET X-ray survey is clear in this panel, where we see that the $w$ sensitivity of the SZE survey peaks near $z \sim 1$ (solid line). For the X-Ray survey, the $w$ sensitivity peaks at $z \sim 0.7$ (dashed line). However, it is clear that the sensitivity per unit cluster ($(dF_{ww}^S/dz)/(dN/dz)$, denoted by dotted line for SZE and dot-dashed line for X-Ray) increases as we go to higher redshift. This emphasizes the importance of high redshift cluster surveys for probing $w$.  High redshift clusters in the DUET survey are even more sensitive to $w$, because they are more massive and lie well beyond the exponential cutoff in the mass function;  however, these clusters are so rare that none are detected in the DUET survey.

The sharp cutoff and first mini--peak at redshift $z\sim0.3$ of $dF_{ww}^S/dz$ for the survey is a direct result of our requirement that clusters have masses above $10^{14}h^{-1}M_\odot$.  In Fig~\ref{fig:surveys} the plot of the limiting mass becomes flat at redshifts below $z\sim0.3$; we include this mass cutoff for several reasons:  (1) the mass--observable scaling relations we've adopted are for clusters, and they are inappropriate for group scale systems and (2) a flux limited survey becomes sensitive to surface brightness limitations at low redshift, where the flux from these nearby objects is spread over a larger and larger solid angle.  Introducing a minimum mass in our survey causes this interesting artifact in the $w$ sensitivity at low redshift, which
can be understood by considering the competitive behaviour of the $w$-dependences of the limiting mass, the survey volume element, and the growth factor of density perturbations. As long as the limiting mass is constant, the opposite sensitivities of the volume element and the growth factor to $w$ determine the net $w$-sensitivity of the survey. At low redshift, the $w$-sensitivity of the volume element dominates over that of the growth factor. Beyond redshift $\sim0.3$, the limiting mass suddenly rises above the minimum, allowing the net $w$ sensitivity to include that of the limiting mass.  The limiting mass is sensitive to $w$ primarily through the angular diameter or luminosity distance to the redshift of interest, and this sensitivity combines with that of the growth factor to offset to a larger degree the $w$ sensitivity of the volume element.  Note that $dF_{ww}^S/dz$ is positive definite  which is the reason for the visual appearance of a break in the sensitivity, which is actually reflecting an underlying change in sign of the $w$ sensitivity.

For the followup (upper panel), the sensitivity per unit cluster peaks at $z \sim 0.8$ for both the surveys. However, survey weighted sensitivities of the followup to variation in $w$ are peaked
at much lower redshifts, where the survey yields are much higher.  This is due to the fact that the $w$-sensitivity of the followup is a balance between  the $w$-sensitivities of the mass observable and the number of clusters one has at that redshift to make the measurement.  The fact that the redshift dependence of the cluster mass observable is similar for both X--ray and SZE just reflects the weak mass dependence of this sensitivity.  Survey strategists should consider a followup program that targets predominantly redshift $z\sim1$ clusters if their goal is to constrain the evolution parameter and improve constraints on $w$; however, having evolution information over the entire redshift range of the survey is critical to testing the form of the non-standard evolution model, which we parametrize here simply as $(1+z)^\gamma$.

\section{Discussion and Conclusions}
\label{sec:conclusions}
Any attempt to precisely measure the dark energy equation of state $w$ with cluster surveys will require (i) a strong external prior on the curvature (presumably from CMB anisotropy studies) and (ii) an understanding of the evolution of the relation between cluster halo mass and observable properties like the X--ray luminosity, SZE luminosity, galaxy light or weak lensing shear.  We have examined the effects of current uncertainties about cluster structural evolution; for two recently proposed cluster surveys the estimated constraints on $w$ are $\sim$2--3 times weaker than if one assumes full knowledge of cluster evolution.  Constraints on other interesting cosmological parameters are also weakened (see Table 1).

Followup observations to measure cluster masses directly will enable one to solve for cluster structure evolution and to enhance cosmological constraints.  We have examined the effects of followup mass observations from hydrostatic or dynamical methods, and we find that even modest followup of 1\% of the cluster sample can improve survey constraints.  Full followup with mass measurements that are 30\% uncertain, on average, provide cosmological constraints that match or surpass those possible through $dN/dz$ alone with full knowledge of cluster evolution.  Full followup with weak lensing mass measurements is currently being planned for the SPT SZE survey.

The implications are quite interesting.  Essentially, to do precision cosmology with cluster surveys and followup, we need only know that clusters conform to mass--observable scaling relations, and that these relations evolve in some well behaved manner.  Then, together with our well established theoretical framework for structure formation, the observed cluster redshift distribution and followup masses of as few as 1\% of the sample then provide enough information to deliver precise constraints on cosmological parameters and the character and evolution of the mass--observable relation simultaneously.  In a sense cluster surveys with limited followup are self--calibrating:  one gains detailed knowledge of the structure of the tracers (i.e. galaxy clusters) and  detailed knowledge of the evolution of the universe from the same dataset.

We have focused here on the mean equation of state parameter $w$, and for the two surveys considered, we examine the redshift variation of the sensitivity to $w$.  In the SPT SZE survey,
the sensitivity to the dark energy equation of state peaks at $z\sim1$ with sensitivity at or above half the peak for $0.65\le z\le1.5$.  In the DUET X-ray survey, the sensitivity peaks at $z\sim0.7$ with sensitivity at or above half the peak for $0.45\le z\le 1.0$.  In the case of both surveys the $w$ sensitivity of followup mass measurements peaks near redshift $z\sim0.35$.  In the case of the cluster redshift distribution, the most information about $w$ is provided by the highest redshift clusters, and so deeper, more sensitive surveys will in general be better for studies of the dark energy equation of state.

One interesting feature of our analysis is the orientation of the elliptical constraints on $w$ and $\Omega_M$ (see Fig.~\ref{fig:cosmo}).   In general, the rotation of the parameter degeneracy can be understood as the result of competing effects of changes in the volume element and the growth factor as parameters vary.  Variations in $w$ (and $\Omega_M$) affect the survey yield in different ways at different redshifts, and so the $w$-$\Omega_M$ degeneracy depends on the redshift distribution of a particular survey.  Rotations of parameter degeneracies occur as the maximum redshift of the survey is varied \citep{levine02,hu02}.  We have further found that changing the prior on $\Omega_{tot}$ and changing the degree of mass followup on a survey also result in rotations of the parameter degenaracy.  This behavior has interesting implications for the design of cluster surveys that are optimally complementary to CMB anisotropy and SNe Ia distance measurements, and it deserves further study.

In addition, we emphasize that the final constraint on the determination of cosmological parameters depends sensitively on the survey strategy and also the details of the followup. For example, a different definition of $M_f(\theta)$ would lead to slightly different uncertainties. Changing $\theta$ from a  quantity that varies with redshift to some fixed value leads to modest variations of the constraints. In general, best results can be obtained by optimizing $M_f(\theta)$ so that one probes as mush of the virial mass as possible.

\acknowledgments
SM thanks Ben Wandelt and Shiv Sethi for helpful conversations.  JM thanks Zoltan Haiman for many fun discussions of cluster surveys.  This work has been supported by NASA Long Term Space Astrophysics grant NAG5--11415 and {\it Chandra} X--ray Observatory archival grant AR1--2002X, awarded through the Smithsonian Astrophysical Observatory.

\bibliographystyle{../Bib/Astronat/apj}
\bibliography{../Bib/cosmology}

\end{document}